\documentclass[aps,prd,twocolumn,showpacs,showkeys,superscriptaddress,floatfix,elsart]{revtex4-1}
\usepackage{latexsym}
\usepackage{amssymb}
\usepackage{amsmath}
\usepackage{amscd}
\usepackage{amsthm}
\usepackage{graphicx}
\usepackage{textcomp}
\usepackage{colortbl}
\usepackage{hyperref}
\usepackage[font={footnotesize,it}]{caption}
\usepackage{multirow}
\usepackage[T1]{fontenc}
\usepackage{ae,aecompl}

\begin{document}
\title{Transitioning Universe with hybrid scalar field in Bianchi I space-time}
\author{Anil Kumar Yadav}
\email[]{abanilyadav@yahoo.co.in}
\affiliation{Department of Physics, United College of Engineering and Research, Greater Noida - 201310, India}
\author{A. M. Alshehri}
\email[]{amshehri@kku.edu.sa}
\affiliation{Department of Physics, College of Science, King Khalid University, P.O. Box: 9004, Abha, 61413, Saudi Arabia}
\author{Nafis Ahmad}
\email[]{nafis.jmi@gmail.com}
\affiliation{Department of Physics, College of Science, King Khalid University, P.O. Box: 9004, Abha, 61413, Saudi Arabia}
\author{G. K. Goswami}
\email[]{gk.goswami9@gmail.com}
\affiliation{Department of Mathematics, Netaji Subhas University of Technology, Delhi, India}
\author{Mukesh Kumar}
\email[]{mukesh.kumar@gla.ac.in}
\affiliation{Department of Mathematics, GLA University, Mathura - 281406 India}
\begin{abstract}
In this paper we investigate a Bianchi type I transitioning Universe in Brans-Dicke theory. To get an explicit solution of the field equations, we assume scalar field as $\phi = \phi_{0}\left[t^{\alpha}exp(\beta t)\right]^{\delta}$ with $\phi_{0}$, $\alpha$, $\beta$ and $\delta$ as constants. The values of $\alpha$ and $\beta$ are obtained by probing the proposed model with recent observational Hubble data (OHD) points. The interacting and non-interacting scenarios between dark matter and dark energy of the derived Universe within the framework of Brans-Dicke gravity are investigated. The $om(z)$ analysis of the Universe in derived model shows that the Universe is filled with dynamical dark energy with its equation of state parameter $\omega_{de} > -1$. Hence the Universe behaves like a quintessence model at present epoch. Some physical properties of the Universe are also discussed.
\end{abstract}
\maketitle
\section{Introduction}\label{1}
SN Ia observations \cite{Perlmutter/1999,ref2} followed by CMBR anisotropy spectrum\cite{ref3}, large scale structure (LSS)\cite{ref4} and Planck results for CMB anisotropies \cite{ref5} ascertain that the observable Universe exists in an accelerated expansion phase at present. While in the past, the Universe was in decelerating expnsion phase during radiation and structure formation era. In refs. \cite{ref6,ref7}, the authors have investigated that in the current universe, the critical energy density is large. So, a substantial amount of energy component apart from the baryon matter density is present in the Universe. Moreover, the missing energy (dark energy) should speed up the cosmic expansion of the Universe in order to explain the observed supernovae red shift - magnitude relation. This is possible only when the so called dark energy (DE) have negative pressure to counteract gravitational pressure of barionic matter. The recent observations indicate that DE dominates the Universe at present and it covers nearly $70 \% $ of the total energy contents of the Universe. The cosmological constant $(\Lambda)$ has repulsive character so, $\Lambda$ becomes a natural choice for dark energy but it has very low magnitude as compared to the particle physics expectations. This problem associated with $\Lambda$ is known as ``fine tuning problem''. Another serious issue with $\Lambda$ is that its repulsive character does not support early radiation dominated hot universe which require dominant attractive field to support decoupling. This problem is called ``cosmic coincidence problem''. So, a dynamical cosmological constant $(\Lambda = \Lambda(t))$ with negative pressure looks promising candidate of DE. Carroll and Hoffman \cite{ref8} have investigated a model of accelerating Universe in which DE is considered as a fluid with equation of state (EoS) parameter $\omega_{de} = \frac{p_{de}}{\rho_{de}}$ in a conventional manner. Many researchers \cite{ref9}$-$\cite{ref16} have also constructed cosmological models of the Universe with a variable equation of state (EoS) parameter $\omega_{de}$ of DE and considered different parametric forms of $\omega_{de}$. Note that for cosmological constant dominated Universe, $\omega_{de} = -1$ while $-1\leq\omega_{de} < 0$ and  $\omega_{de} \leq -1 $ corresspond to quintessence model \cite{ref17,ref18} and phantom model\cite{ref19,ref20} of the Universe respectively.\\

Initially, Ernst Mach had been discarded the idea of absolute space and time and proposed that inertial properties occured in the matter are due to its interaction with distant background in the Universe. Impressed with this idea, Einstein had proposed the interconnection between gravitation and curvature. In spite, GTR does not fully satisfy Mach's conjecture \cite{ref41}. Then Jordon followed by Brans and Dicke had investigated the modified relativistic theories of gravitation which satisfy Mach's principle \cite{Brans/1961,ref44}. Brans - Dicke (BD) theory is a scalar tensor theory which has a constant coupling parameter $\omega$ and scalar field $\phi$. This theory acts as interaction of local matter with the distant background. We recover GTR from BD theory when $\omega$ $\to$ $\infty$ \cite{ref41}. The solar system tests \cite{ref45}, Viking Space  Probe \cite{ref46} and the recent experimental evidences \cite{ref47, ref48} put lower limit on coupling constant $\omega$ higher than 40000. The pioneering studies with BD scalar field, in particular, evolution of the Universe, inflation, structure formation, early as well as late time behaviors of the Universe, quintessences and high energy description of dark energy in an approximate 3-brane are given in refs.  \cite{ref49}-\cite{ref73}.\\

In the present Universe, it is reasonable to consider the interaction between dark components: dark matter (DM) and DE. In facts these scenarios resolve the problems associated with $\Lambda$ and provide an instinctive way to detect DE \cite{Cimento/2004,Setare/2007}. It is worthwhile to note the results of some observation are in favor of such type of interaction \cite{Bertolami/2007,Delliou/2007,Berger/2006,Valentino/2017}. Since the actual nature of DM and DE are still unknown therefore one can not exclude the interaction between DM and DE. Some important applications of such type of interactions are given in Refs. \cite{Tocchini/2002,Farrar/2004,Guo/2007,Kumar/2019,Hassan/2020a}. Furthermore, it has been proved that when an interaction between holographic dark energy (HDE) and DM is taken into account, the phantom line is crossed in the BD cosmology \cite{Jamil/2012}. In Chimento \cite{Chimento/2010}, linear and nonlinear interactions in the dark sector of the Universe have been studied with energy transfer and shown that they can be considered as unified model. Motivated by the above ideas, in this paper, we confine ourselves to study the possible interaction between pressure less DM and DE with hybrid scalar field in BD cosmology. In Brans-Dicke theory, there is an additional wave equation for the BD scalar field, $\frac{\ddot{\phi}}{\phi}+3\frac{\dot{a}\dot{\phi}}{a\phi} = \frac{T}{(2\omega+3)}$. Here, T stands for trace of the stress energy tensor of dark matter and dark energy.\\

From astrophysical observations, it has been confirmed now that the neutrinos and Cosmic Microwave Background Radiations(CMBR) are left over parts of the primordial fireball. The neutrinos and CMBR are $0.1$ to $0.5$ \%  and $0.01$ \% of the total matter or energy content of the Universe. Neutrino viscosity creates anisotropy in the Universe which dissipates out with the advent of time \cite{ref21,ref22}. The Willkinsion Microwave Anisotropic Probe (WMAP) \cite{ref23} also creates interest in the investigation of non-isotropic models of the Universe. Accordingly, a large
number of spatially homogeneous and anisotropic solutions of Einstein's field equations in General Theory of Relativity (GTR) have been investigated \cite{ref24}-\cite{Amirhashchi/2020}. In fact in 1962, Spatially homogeneous and anisotropic cosmology was formulated by Heckmann and Schucking \cite{ref40}. This is the reason for spatially homogeneous and anisotropic Bianchi type-I metric to be referred as Heckmann-Schuking metric. Some useful applications of BD scalar field in Bianchi I space time are given in Refs. \cite{ref71,Sharma/2018}.\\

In this paper, we investigate a Bianchi type I transitioning universe in Brans-Dicke gravity. To get an explicit solution of the field equations, we assume scalar field as $\phi = \phi_{0}\left[t^{\alpha}exp(\beta t)\right]^{\delta}$ with $\phi_{0}$, $\alpha$, $\beta$ and $\delta$ as constants. The values of model parameters $\alpha$ and $\beta$ are obtained by constraining the Universe in derived model with recent observational Hubble data sets. The interacting and non-interacting scenarios between dark matter and dark energy of transitioning Universe within the framework of Brans-Dicke gravity are investigated. The paper is structured as follows: in section \ref{2}, we have given the basic mathematical formalism of the model. Section \ref{3} deals with the observational confrontation of model with recent $H(z)$ data. The non-interacting and interacting scenarios of the Universe in dervied model are discussed in sections \ref{4} and \ref{5}. We have presented discussions on the physical properties of the model in sec. \ref{6}. In the last section \ref{7}, we have summarized our findings.\\
\section{The model and Basic equations}\label{2}
The Brans-Dicke action in Jordan frame is read as
\begin{equation}
\label{action-1}
S = \frac{1}{16\pi }\int\sqrt{-g}\left[\phi R-\omega \frac{\phi_{, i}\phi^{, i}}{\phi}+L_{m}\right]d^{4}x
\end{equation} 
where $\omega$, $\phi$ and $L_{m}$ denote the Brans-Dicke coupling constant, Brans-Dicke scalar field and Lagrangian for all matter field respectively. If one consider the matter field to be consist of dark matter (DM) and dark energy (DE) then the field equations in Brans-Dicke theory \cite{Brans/1961} is given by
\[
R_{ij}-\frac{1}{2}Rg_{ij}=-\frac{8\pi}{\phi }(T_{ij}^{m}+T_{ij}^{de})
\]
\begin{equation}
\label{BD-1}
-\frac{\omega}{\phi^{2}}\left(\phi_{i}\phi_{j}-\frac{1}{2}g_{ij}\phi_{k}\phi^{k}\right)
-\frac{1}{\phi}(\phi_{ij}-g_{ij}\square\phi)
\end{equation}
and
\begin{equation}
\label{BD-2}
\square\phi=\frac{8\pi (T^{m}+T^{de})}{(2\omega+3)}
\end{equation}
where $T_{ij}^{m}$ and $T_{ij}^{de}$ are the energy momentum tensor for DM and dark DE respectively.\\
The energy conservation equation is read as
\begin{equation}
\label{ec}
(T^{m}_{ij}+T_{ij}^{de});j = 0
\end{equation}
Equation (\ref{ec}) leads to
\begin{equation}
\label{ec-1}
\dot{\rho_{de}}+\dot{\rho_{m}}+3H(1+\omega_{de})\rho_{de}+3H\rho_{m} = 0
\end{equation}
where $\rho_{m}$ and $\rho_{de}$ are the energy density of DM and DE respectively. $\omega_{de} = p_{de}/\rho_{de}$ is the equation of state parameter of dark energy and over dot denotes derivatives with respect to time $(t)$. $H$ is the Hubble's function and it is defined as $H = \frac{\dot{a}}{a}$ with $a$ as the average scale factor. The detail discription of field equations and energy conservation equation of Brans-Dicke theory are given in Ref. \cite{Aditya/2019}. It is worthwhile to mention that energy conservation equation and related issues in modified theories of gravity and BD cosmology have been described in Refs. \cite{Koivisto/2006,Amirhashchi/2019,Narlikar/2002} (see Appendix). Moreover, the violation of strong energy condition (SEC) is expected in accelerating Universe. In Refs. \cite{Sen/2008,Qiu/2008}, the validation/violation of energy conditions have been checked by using observational data.\\

The Bianch type I space-time is read as
\begin{equation}
\label{BI}
ds^{2} = dt^{2}-A^{2}(t)dx^{2}-B(t)^{2}dy^{2}-C(t)^{2}dz^{2})
\end{equation}
where $A(t),~B(t)~\& C(t)$ are scale factors along $x$, $y$ and $z$ direction respectively and average scale factor is defined as $a = (ABC)^{\frac{1}{3}}$.\\
The field equations (\ref{BD-1}) for space-time (\ref{BI}) are obtained as
\begin{equation}
\label{ef-1}
\frac{\ddot{B}}{B}+\frac{\ddot{C}}{C}+\frac{\dot{B}\dot{C}}{BC}+\frac{\omega}{2}\frac{\dot{\phi}^{2}}{\phi^{2}}+\frac{\dot{\phi}}{\phi}\left(\frac{\dot{B}}{B}+\frac{\dot{C}}{C}\right)+\frac{\ddot{\phi}}{\phi}=-\frac{\omega_{de}\rho_{de}}{\phi}
\end{equation}
\begin{equation}
\label{ef-2}
\frac{\ddot{A}}{A}+\frac{\ddot{C}}{C}+\frac{\dot{A}\dot{C}}{AC}+\frac{\omega}{2}\frac{\dot{\phi}^{2}}{\phi^{2}}+\frac{\dot{\phi}}{\phi}\left(\frac{\dot{A}}{A}+\frac{\dot{C}}{C}\right)+\frac{\ddot{\phi}}{\phi}=-\frac{\omega_{de}\rho_{de}}{\phi}
\end{equation}
\begin{equation}
\label{ef-3}
\frac{\ddot{A}}{A}+\frac{\ddot{B}}{B}+\frac{\dot{A}\dot{B}}{AB}+\frac{\omega}{2}\frac{\dot{\phi}^{2}}{\phi^{2}}+\frac{\dot{\phi}}{\phi}\left(\frac{\dot{A}}{A}+\frac{\dot{B}}{B}\right)+\frac{\ddot{\phi}}{\phi}=-\frac{\omega_{de}\rho_{de}}{\phi}
\end{equation}
\begin{equation}
\label{ef-4}
\frac{\dot{A}\dot{B}}{AB}+\frac{\dot{B}\dot{C}}{BC}+\frac{\dot{A}\dot{C}}{AC}-\frac{\omega}{2}\frac{\dot{\phi}^{2}}{\phi^{2}}+\frac{\dot{\phi}}{\phi}\left(\frac{\dot{A}}{A}+\frac{\dot{B}}{B}+\frac{\dot{C}}{C}\right) = \frac{\rho_{de}+\rho_{m}}{\phi}
\end{equation}
\begin{equation}
\label{ef-5}
\frac{\ddot{\phi}}{\phi}+\left(\frac{\dot{A}}{A}+\frac{\dot{B}}{B}+\frac{\dot{C}}{C}\right)\frac{\dot{\phi}}{\phi}
= \frac{\rho_{de}(1-3\omega_{de})+\rho_{m}}{3+2\omega}
\end{equation}
Solving equations(\ref{ef-1})-(\ref{ef-3}), we have following system of equations
\begin{equation}
\label{ef-7}
\frac{\ddot{A}}{A}-\frac{\ddot{B}}{B}+\frac{\dot{A}\dot{C}}{AC}-\frac{\dot{B}\dot{C}}{BC}+\left(\frac{\dot{A}}{A}-\frac{\dot{B}}{B}\right)\frac{\dot{\phi}}{\phi} = 0
\end{equation}
\begin{equation}
\label{ef-8}
\frac{\ddot{B}}{B}-\frac{\ddot{C}}{C}+\frac{\dot{A}\dot{B}}{AB}-\frac{\dot{A}\dot{C}}{AC}+\left(\frac{\dot{B}}{B}-\frac{\dot{C}}{C}\right)\frac{\dot{\phi}}{\phi} = 0
\end{equation}
\begin{equation}
\label{ef-9}
\frac{\ddot{C}}{C}-\frac{\ddot{A}}{A}+\frac{\dot{B}\dot{C}}{BC}-\frac{\dot{A}\dot{B}}{AB}+\left(\frac{\dot{C}}{C}-\frac{\dot{A}}{A}\right)\frac{\dot{\phi}}{\phi} = 0
\end{equation}
Subtracting Eq. (\ref{ef-9}) from Eq. (\ref{ef-7}), we obtain
\begin{equation}
\label{n-1}
2\frac{\ddot{A}}{A}+\frac{\dot{A}}{A}\left(\frac{\dot{B}}{B}+\frac{\dot{C}}{C}\right)+\left(2\frac{\dot{A}}{A}-\frac{\dot{B}}{B}-\frac{\dot{C}}{C}\right)\frac{\dot{\phi}}{\phi} = \frac{\ddot{B}}{B}+\frac{\ddot{C}}{C}+2\frac{\dot{B}\dot{C}}{BC}
\end{equation}
On integrating Eq. (\ref{n-1}), we obtain
\begin{equation}
\label{n-2}
\frac{\dot{B}}{B}+\frac{\dot{C}}{C}-2\frac{\dot{A}}{A} = \frac{\Gamma}{ABC\phi}
\end{equation}
where $\Gamma$ is constant of integration which can be evaluated by applying initial condition. It is well known that the Universe has either singular or non singular origin. For singular Universe which has a point type big bang singularity at $t = 0$, we have $A =0$, $B = 0$ and $C = 0$ at initial epoch $t = 0$. Therefore, from equation (\ref{n-2}), we obtain $\Gamma =0$. Again for non-singular Universe, the directional scale factors are constant at initial moment which also leads $\Gamma =0$.\\
Thus, Eq. (\ref{n-2}) leads to
\begin{equation}
\label{n-3}
\frac{\dot{B}}{B}+\frac{\dot{C}}{C}-2\frac{\dot{A}}{A} = 0
\end{equation}
Integrating Eq. (\ref{n-3}), we obtain 
\begin{equation}
\label{s-1}
A^{2} = BC \Rightarrow B = AD \;\;\; \& \;\; C = \frac{A}{D}
\end{equation}
where $D = D(t)$ measures the anisotropy in Bianchi type I universe.\\

Eqs. (\ref{ef-8}) and (\ref{s-1}) lead to
\[
\frac{\ddot{D}}{D}-\frac{\dot{D}^{2}}{D^{2}}+\frac{\dot{D}}{D}\left(3\frac{\dot{A}}{A}+\frac{\dot{\phi}}{\phi}\right) = 0
\]
\begin{equation}
\label{s-2}
\Rightarrow \; \frac{\frac{d}{dt}\left(\frac{\dot{D}}{D}\right)}{\left(\frac{\dot{D}}{D}\right)} = - \left(3\frac{\dot{A}}{A}+\frac{\dot{\phi}}{\phi}\right)
\end{equation}
After integration of equation (\ref{s-2}), we obtain
\begin{equation}
\label{s-3}
D = exp\left[\int\frac{k}{A^{3}\phi}dt\right]
\end{equation}
where $k$ is the constant of integration.\\
Now, the average scale factor is read as
\begin{equation}
\label{s-2}
a^{3} = ABC = A^{3} \Rightarrow a = A
\end{equation}

Finally, we obtain the following equations which yet to be solve to describe the feature of proposed model
\begin{equation}
\label{ef-H1}
2\frac{\ddot{a}}{a}+\frac{\dot{a}^{2}}{a^{2}}+\frac{k^{2}}{a^{6}\phi^{2}}+\frac{\omega}{2}\frac{\dot{\phi}^{2}}{\phi^{2}}+2\frac{\dot{\phi}}{\phi}\frac{\dot{a}}{a} + \frac{\ddot{\phi}}{\phi} = -\frac{\omega_{de}\rho_{de}}{\phi}
\end{equation}
\begin{equation}
\label{ef-H2}
3\frac{\dot{a}^{2}}{a^{2}}-\frac{k^{2}}{a^{6}\phi^{2}}-\frac{\omega}{2}\frac{\dot{\phi}^{2}}{\phi^{2}}+3\frac{\dot{\phi}}{\phi}\frac{\dot{a}}{a} =  \frac{\rho_{de}+\rho_{m}}{\phi}
\end{equation}
\begin{equation}
\label{scalarfield}
\frac{\ddot{\phi}}{\phi}+3\frac{\dot{a}\dot{\phi}}{a\phi}=\frac{\rho_{de}(1-3\omega_{de})+\rho_{m}}{3+2\omega}
\end{equation}
Following, Johri and Desikan and references therein \cite{Johari/1994,Johari/1989,Ali/2014,Singh/2012,Sheykhi/2009}, we consider following relation between Brans-dicke scalar field $\phi$ and average scale factor $a$ as
\begin{equation}
\label{bd-1}
\phi=\phi_{0}a^{\delta}
\end{equation}
where $\phi_{0}$ and $\delta$ are constants. In principle, there is no compelling reason for the
choice of $\phi=\phi_{0}a^{\delta}$. However, we shall see in due course that for small $\mid \delta \mid$, the choice (\ref{bd-1}) leads to consistent results (see Ref. \cite{Banerjee/2007} for detail).\\

The type Ia supernovae observations \cite{Perlmutter/1998,Perlmutter/1999,Riess/2004,Tonry/2003}, CMB anisotropies \cite{Bennett/2003} and recently Plank Collaborations \cite{Aghanim/2018} have confirmed that the present Universe is in accelerating phase but it was in decelerating phase in past. Therefore, the Universe must have a signature flipping from past decelerated expansion to current accelerated expansion \cite{Padmanabhan/2003,Amendola/2003}. So, to obtain an exact model of transitioning Universe, we have to assume scale factor in the from of $a=t^{\alpha}exp(\beta t)$ with $\alpha$ and $\beta$ as arbitrary constants. In the literature, such type of ansatz for scale factor is referred as hybrid expansion law for evolution of Universe and it gives a time varying deceleration parameter
\cite{Akarsu/2014,Yadav/2012,Yadav/2016}.\\
Therefore, equation (\ref{bd-1}) is recast as
\begin{equation}
\label{bd-2}
\phi = \phi_{0}\left[t^{\alpha}exp(\beta t)\right]^{\delta}
\end{equation}
\begin{figure}\label{fig2}
\begin{center}
\includegraphics[width=7cm]{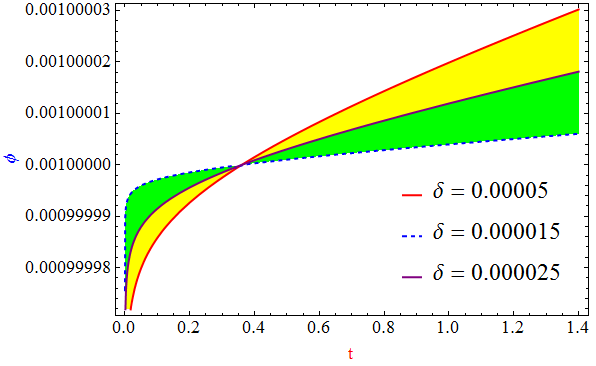}
\end{center}
\caption{The plot of scalar field $\phi$ versus time $t$. The unit of time is Gyrs.}
\end{figure}
The physical behavior of scalar field $\phi$ with respect to $t$ for different values of $\delta$ is shown in Fig. 1. We observe that the scalar field increases with passage of time.
\section{Observational confrontation}\label{3}
\begin{figure}\label{fig1}
\begin{center}
\includegraphics[width=8.5cm,height=7.5cm]{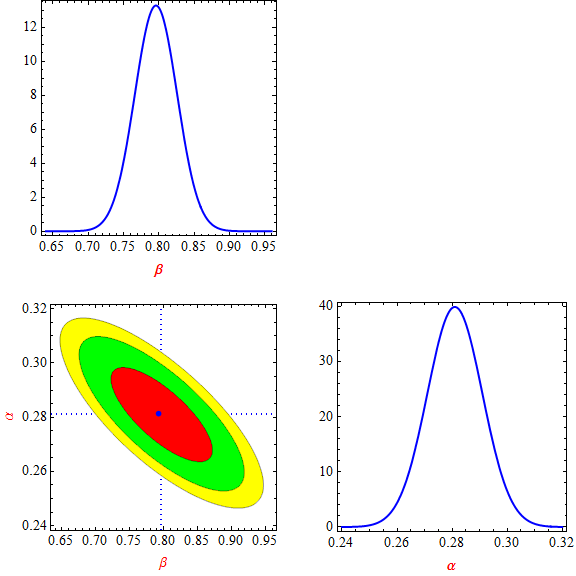}
\end{center}
\caption{One dimensional marginalized distribution and two dimensional contours with 68.3\% (inner contour), 95.4\% (middle contour) and 99.7\% (outer contour) confidence level (CL) around the best fit values as $\alpha = 0.281 \pm 0.004 $ \& $ \beta = 0.796 \pm 0.012$ in the $\beta - \alpha$ plane obtained by fitting the derived model with 46 OHD points.}
\end{figure}
In this section, we find constraints on the model parameters $\alpha$ and $\beta$ by bounding the model under consideration with recent $46$ observational Hubble data (OHD) points in the red-shift range $0 \leq z \leq 2.36$ with their corresponding standard deviation $\sigma_{i}$. These $46$ OHD points are compiled in Table 1 of this paper and in Refs. \cite{Prasad/2020,Amirhashchi/2019,Farooq/2017}. The cosmic chronometric (CC) technique is adopted to determined these uncorrelated data. There reason behind to take this data is the fact that OHD data obtained from
CC technique is model-independent. In fact, the most evolving galaxies based on the ``galaxy differential age'' method is used to determine this CC data. Here, we adopt the present value of Hubble constant as $H_{0} = 67.8\pm 1$ km/s/Mpc as observed by Plank Collaboration \cite{ref5}.
\begin{table*}
\begin{center}
\small
\caption{Hubble parameter H(z) with redshift and errors $\sigma_{i}$.\label{tbl-1}}
\begin{tabular}{@{}crrrrrrrrrrr@{}}
\hline
S.N.~~~~&~~~~z~~~~~& ~~~~H(z)~~~~ & $\sigma_{i}$~~~~&~~~~~~Method~~~~ &~~~~~~ References \\
\hline
1~~~~&~~~~0~~~~~ &~~~ 67.77~~~ & ~~~ 1.30~~~~ & ~~~~~ DA ~~~~&~~~~\cite{Macaulay/2018} \\
2~~~~&~~~~~0.07~~~~ &~~~~~ 69~~~~ & ~~~~~ 19.6~~~~ & ~~~~~ DA ~~~~&~~~~ \cite{Zhang/2014}\\
3~~~~&~~~~~0.09~~~~ &~~~~~ 69~~~~ & ~~~~~ 12~~~~ & ~~~~~ DA ~~~~&~~~~ \cite{Simon/2005} \\
4~~~~&~~~~~0.01~~~~ &~~~~~ 69~~~~ & ~~~~~ 12~~~~ & ~~~~~  DA ~~~~&~~~~\cite{Stern/2010} \\
5~~~~&~~~~~0.12~~~~ &~~~~~ 68.6~~~~ & ~~~~~ 26.2~~~~ & ~~~~~  DA ~~~~&~~~~ \cite{Zhang/2014}\\
6~~~~&~~~~~0.17~~~~ &~~~~~ 83~~~~ & ~~~~~ 8~~~~ & ~~~~~ DA ~~~~&~~~~\cite{Stern/2010} \\
7~~~~&~~~~~0.179~~~~ &~~~~~75~~~~ & ~~~~~ 4~~~~ & ~~~~~  DA ~~~~&~~~~\cite{Moresco/2012} \\
8~~~~&~~~~~0.1993~~~~ &~~~~~75~~~~ & ~~~~~ 5~~~~ & ~~~~~  DA ~~~~&~~~~\cite{Moresco/2012} \\
9~~~~&~~~~~0.2~~~~ &~~~~~ 72.9~~~~ & ~~~~~ 29.6~~~~ & ~~~~~  DA ~~~~&~~~~\cite{Zhang/2014}\\
10~~~~&~~~~~0.24~~~~ &~~~~~ 79.7~~~~ & ~~~~~ 2.7~~~~ & ~~~~~  DA ~~~~&~~~~\cite{Gazta/2009}\\
11~~~~&~~~~~0.27~~~~ &~~~~~ 77~~~~ & ~~~~~ 14~~~~ & ~~~~~  DA ~~~~&~~~~\cite{Stern/2010} \\
12~~~~&~~~~~0.28~~~~ &~~~~~ 88.8~~~~ & ~~~~~ 36.6~~~~ & ~~~~~  DA ~~~~&~~~~\cite{Zhang/2014}\\
13~~~~&~~~~~0.35~~~~ &~~~~~ 82.7~~~~ & ~~~~~ 8.4~~~~ & ~~~~~  DA ~~~~&~~~~\cite{Chuang/2013}\\
14~~~~&~~~~~0.352~~~~ &~~~~~83~~~~ & ~~~~~ 14~~~~ & ~~~~~  DA ~~~~&~~~~\cite{Moresco/2012} \\
15~~~~&~~~~~0.38~~~~ &~~~~~81.5~~~~ & ~~~~~ 1.9~~~~ & ~~~~~  DA ~~~~&~~~~\cite{Alam/2016} \\
16~~~~&~~~~~0.3802~~~~ &~~~~~83~~~~ & ~~~~~ 13.5~~~~ & ~~~~~  DA ~~~~&~~~~\cite{Moresco/2016} \\
17~~~~&~~~~~0.4~~~~ &~~~~~ 95~~~~ & ~~~~~ 17~~~~ & ~~~~~  DA ~~~~&~~~~\cite{Simon/2005} \\
18~~~~&~~~~~0.4004~~~~ &~~~~~77~~~~ & ~~~~~ 10.2~~~~ & ~~~~~  DA ~~~~&~~~~\cite{Moresco/2016} \\
19~~~~&~~~~~0.4247~~~~ &~~~~~87.1~~~~ & ~~~~~ 11.2~~~~ & ~~~~~  DA ~~~~&~~~~\cite{Moresco/2016} \\
20~~~~&~~~~~0.43~~~~ &~~~~~ 86.5~~~~ & ~~~~~ 3.7~~~~ & ~~~~~  DA ~~~~&~~~~\cite{Gazta/2009}\\
21~~~~&~~~~~0.44~~~~ &~~~~~ 82.6~~~~ & ~~~~~ 7.8~~~~ & ~~~~~  DA ~~~~&~~~~\cite{Blake/2012}\\
22~~~~&~~~~~0.44497~~~~ &~~~~~ 92.8~~~~ & ~~~~~ 12.9~~~~ & ~~~~~  DA ~~~~&~~~~\cite{Moresco/2016}\\
23~~~~&~~~~~0.47~~~~ &~~~~~ 89~~~~ & ~~~~~ 49.6~~~~ & ~~~~~  DA ~~~~&~~~~\cite{Ratsimbazafy/2017}\\
24~~~~&~~~~~0.4783~~~~ &~~~~~80.9~~~~ & ~~~~~ 9~~~~ & ~~~~~  DA ~~~~&~~~~\cite{Moresco/2016} \\
25~~~~&~~~~~0.48~~~~ &~~~~~ 97~~~~ & ~~~~~ 60~~~~ & ~~~~~  DA ~~~~&~~~~\cite{Stern/2010} \\
26~~~~&~~~~~0.51~~~~ &~~~~~90.4~~~~ & ~~~~~ 1.9~~~~ & ~~~~~  DA ~~~~&~~~~\cite{Alam/2016} \\
27~~~~&~~~~~0.57~~~~ &~~~~~96.8~~~~ & ~~~~~ 3.4~~~~ & ~~~~~  DA ~~~~&~~~~\cite{Anderson/2014} \\
28~~~~&~~~~~0.593~~~~ &~~~~~104~~~~ & ~~~~~ 13~~~~ & ~~~~~  DA ~~~~&~~~~\cite{Moresco/2012} \\
29~~~~&~~~~~0.6~~~~ &~~~~~ 87.9~~~~ & ~~~~~ 6.1~~~~ & ~~~~~  DA ~~~~&~~~~\cite{Blake/2012}\\
30~~~~&~~~~~0.61~~~~ &~~~~~97.3~~~~ & ~~~~~ 2.1~~~~ & ~~~~~  DA ~~~~&~~~~\cite{Alam/2016} \\
31~~~~&~~~~~0.68~~~~ &~~~~~92~~~~ & ~~~~~ 8~~~~ & ~~~~~ DA ~~~~&~~~~\cite{Moresco/2012} \\
32~~~~&~~~~~0.73~~~~ &~~~~~ 97.3~~~~ & ~~~~~ 7~~~~ & ~~~~~  DA ~~~~&~~~~\cite{Blake/2012}\\
33~~~~&~~~~~0.781~~~~ &~~~~~105~~~~ & ~~~~~ 12~~~~ & ~~~~~  DA ~~~~&~~~~\cite{Moresco/2012} \\
34~~~~&~~~~~0.875~~~~ &~~~~~125~~~~ & ~~~~~ 17~~~~ & ~~~~~  DA ~~~~&~~~~\cite{Moresco/2012} \\
35~~~~&~~~~~0.88~~~~ &~~~~~ 90~~~~ & ~~~~~ 40~~~~ & ~~~~~ DA ~~~~&~~~~\cite{Stern/2010} \\
36~~~~&~~~~~0.9~~~~ &~~~~~ 117~~~~ & ~~~~~ 23~~~~ & ~~~~~  DA ~~~~&~~~~\cite{Stern/2010} \\
37~~~~&~~~~~1.037~~~~ &~~~~~154~~~~ & ~~~~~ 20~~~~ & ~~~~~  DA ~~~~&~~~~\cite{Moresco/2012} \\
38~~~~&~~~~~1.3~~~~ &~~~~~ 168~~~~ & ~~~~~ 17~~~~ & ~~~~~  DA ~~~~&~~~~\cite{Stern/2010} \\
39~~~~&~~~~~1.363~~~~ &~~~~~160~~~~ & ~~~~~ 33.6~~~~ & ~~~~~  DA ~~~~&~~~~\cite{Moresco/2015} \\
40~~~~&~~~~~1.43~~~~ &~~~~~ 177~~~~ & ~~~~~ 18~~~~ & ~~~~~  DA ~~~~&~~~~\cite{Stern/2010} \\
41~~~~&~~~~~1.53~~~~ &~~~~~ 140~~~~ & ~~~~~ 14~~~~ & ~~~~~  DA ~~~~&~~~~\cite{Stern/2010} \\
42~~~~&~~~~~1.75~~~~ &~~~~~ 202~~~~ & ~~~~~ 40~~~~ & ~~~~~  DA ~~~~&~~~~\cite{Stern/2010} \\
43~~~~&~~~~~1.965~~~~ &~~~~~186.5~~~~ & ~~~~~ 50.4~~~~ & ~~~~~  DA ~~~~&~~~~\cite{Moresco/2015} \\
44~~~~&~~~~~2.3~~~~ &~~~~~ 224~~~~ & ~~~~~ 8~~~~ & ~~~~~ DA ~~~~&~~~~\cite{Busca/2013} \\
45~~~~&~~~~~2.34~~~~ &~~~~~ 222~~~~ & ~~~~~ 7~~~~ & ~~~~~  DA ~~~~&~~~~\cite{Delubac/2015} \\
46~~~~&~~~~~2.36~~~~ &~~~~~226~~~~ & ~~~~~ 8~~~~ & ~~~~~  DA ~~~~&~~~~\cite{Ribera/2014} \\
\hline
\end{tabular}
\end{center}
\end{table*}

The Hubble's parameter and deceleration parameter are obtained as
\begin{equation}
\label{H-1}
H = \frac{\dot{a}}{a}=\frac{\alpha}{t}+\beta
\end{equation}
\begin{equation}
\label{q-1}
q =-1-\frac{\dot{H}}{H^{2}} = -1+\frac{\alpha}{(\alpha+\beta t)^{2}}
\end{equation}
In order to confront our model with observational data, it is convenient to rewrite Hubble's parameter in terms of $z$. For this sake, we use $a = \frac{a_{0}}{1+z}$. Since the present value of scale factor is $a_{0} = 1$ hence $a = \frac{1}{1+z}$.\\
The time-redshift relation is obtained as
\begin{equation}
\label{t-z}
t=\frac{\alpha  \Upsilon\left(\frac{\beta  \left(\frac{1}{z+1}\right)^{1/\alpha }}{\alpha }\right)}{\beta }
\end{equation}
Where $\Upsilon$ denotes the Lambert function or Product Logarithm. For the sake of simplicity, we assume $f(z) = \Upsilon\left(\frac{\beta  \left(\frac{1}{z+1}\right)^{1/\alpha }}{\alpha }\right)$.\\
Thus the expression of Hubble's parameter terms of $z$ is read as
\begin{equation}
\label{H-2}
H(z) = \beta\left(\frac{1}{f(z)}+1\right)
\end{equation}
\begin{figure}\label{fig2}
\begin{center}
\includegraphics[width=7cm]{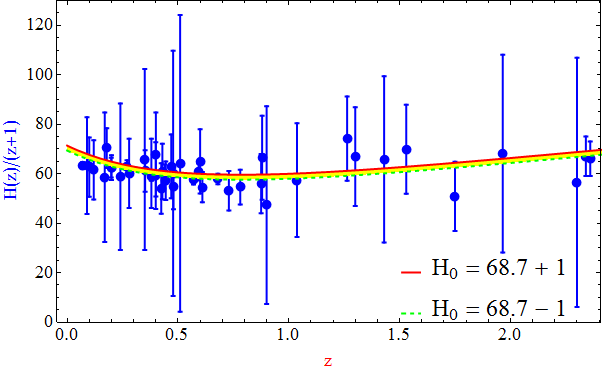}
\end{center}
\caption{The plot of Hubble rate against redshift $z$. The points with error bars indicate the experimental Observation H(z) data. $H_{0}$ is the present value of Hubble constant.}
\end{figure}
\begin{figure}\label{fig2}
\begin{center}
\includegraphics[width=2.7cm]{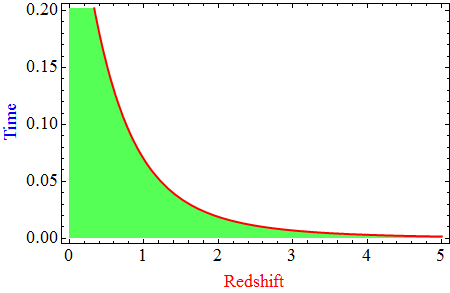}
\includegraphics[width=2.7cm]{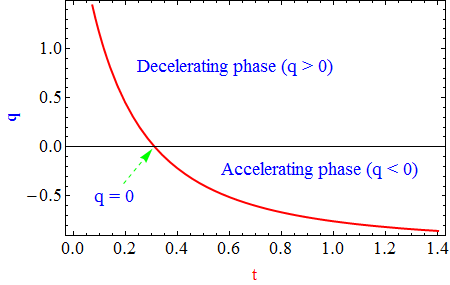}
\includegraphics[width=2.7cm]{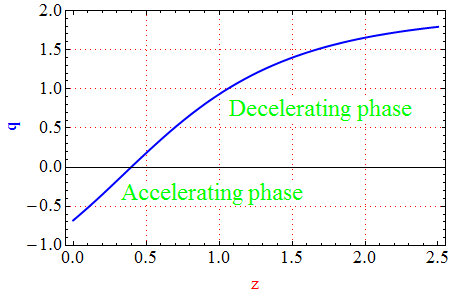}
\end{center}
\caption{The plot of redshift versus time (left panel), variation of deceleration parameter $q$ versus time $t$ (middle panel) and the behavior of deceleration parameter $q$ versus redshift $z$ (right panel) for $\alpha = 0.281$ and $\beta = 0.796$.}
\end{figure}
For constraining model parameters $\alpha$ and $\beta$, we have defined $\chi^{2}$ for parameters with the likelihood given by $\Psi \propto e^{-\frac{\chi^2}{2}}$. Therefore, the $\chi^{2}$ function for OHD points is given as
\begin{equation}
\label{chi-1}
\chi^{2}_{ OHD}=\sum_{i=1}^{N}\left[\frac{H(z_{i},s)-H_{obs}(z_{i})}{\sigma_{i}}\right]^{2}
\end{equation}
where $N = 46$. $s$ and $\sigma_{i}$ represent parameter vector and standard deviation in experimental values of Hubble's function $H$ respectively. \\
The one dimensional marginalized distribution and two dimensional contours with 68.3\%, 95.4\% and 99.7\% confidence level are obtained for our model as shown in Fig. 2. The best fit values of $\alpha$ and $\beta$ come out to be $\alpha = 0.281\pm 0.004$ and $\beta = 0.796\pm 0.012$ at $68.3\%$ CL in the $\beta - \alpha$ plane with $\chi^{2}_{min} = 24.178$. Fig. 3 depicts the plot of Hubble rate $H/(1+z)$ vesus redshift $z$ of derived model for $\alpha = 0.281\pm 0.004$ and $\beta = 0.796\pm 0.012$. From Fig. 4, we observe that the nice fit of derived model with $H(z)$ observational data points. In Fig. 4, we have graphed the dynamical behavior of deceleration parameter $q$ versus time and redshift. From Fig. 4, we conclude that the early Universe was expanding with positive deceleration parameter while the current Universe is in accelerated phase of expansion. Note that for all plots, time is in the unit of Gyrs.
\section{Non interacting model}\label{4}
For non interacting model, we assume that DM and DE interact only gravitationally so that the continuity equation is satisfied separately by each source, and Eq. (\ref{ec}) leads to
\begin{equation}
\label{ec-1}
\dot{\rho_{m}}+3\rho_{m}H =0
\end{equation}
\begin{equation}
\label{ec-2}
\dot{\rho_{de}}+3\rho_{de}(1+\omega_{de})H = 0
\end{equation}
From Eq. (\ref{ec-1}), energy density of DM is obtained as
\begin{equation}
\label{dm-1}
\rho_{m} = \rho_{0}a^{-3} = \rho_{0}\left[t^{\alpha}exp(\beta t)\right]^{-3}
\end{equation}
Using Eqs. (\ref{bd-2}) and (\ref{dm-1}) in Eq. (\ref{ef-H2}), the energy density of DE is obtained as
\begin{widetext}
\begin{equation}
\label{de-1}
\rho_{de} =	{\phi _0 (\delta  (3-\frac{\delta  \omega}{2})+3) (\frac{\alpha}{t} +\beta)^2 \left(t^{\alpha } e^{\beta  t}\right)^{\delta }}-\frac{k^2}{\phi _0\left( t^{\alpha } e^{\beta  t}\right)^{\delta +6}}-\rho _0 t^{-3 \alpha } e^{-3 \beta  t}
\end{equation}
\end{widetext}
Putting the values of $\phi$ and $\rho_{de}$ in Eq. (\ref{ef-1}), we obtain the following expression for equation of state parameter of DE
\begin{widetext}
\begin{equation}
\label{omega}
\omega_{de} =-	\frac{\frac{k^2}{\left(t^{\alpha } e^{\beta  t}\right)^{\delta +6}}+\phi _0^2 \left(t^{\alpha } e^{\beta  t}\right)^{\delta } \left(\left(\delta ^2 \left(\frac{\omega }{2}+1\right)+2 \delta +3\right) (\frac{\alpha}{t} +\beta)^2-\frac{\alpha  (\delta +2)}{t^2}\right)}{\phi^2 _0 (\delta  (3-\frac{\delta  \omega}{2})+3) (\frac{\alpha}{t} +\beta)^2 \left(t^{\alpha } e^{\beta  t}\right)^{\delta }-\frac{k^2}{\left( t^{\alpha } e^{\beta  t}\right)^{\delta +6}}-\phi _0\rho _0 t^{-3 \alpha } e^{-3 \beta  t}}
\end{equation}
\end{widetext}
\section{Interacting model}\label{5}
In this section, we describe the interacting scenario of the Universe by assuming interaction between DM and DE components. Hence the continuity equations for dark matter and dark energy are read as
\begin{equation}
\label{ec-3}
\dot{\rho_{m}}+3\rho_{m}H =Q
\end{equation}
\begin{equation}
\label{ec-4}
\dot{\rho_{de}}+3\rho_{de}(1+\omega_{de})H = -Q
\end{equation}
where $Q$ denotes the coupling between DM and DE. We quantify the coupling between DM and DE is proportional to $\rho_{m}$ and $H$ $i. e.$ $Q \propto \rho_{m}H$. Therefore, for interacting model, we consider $Q = \xi_{1}\rho_{m}H$ with $\xi_{1}$ as constant.

Thus, the expressions for energy densities of DM and DE are respectively given by
\begin{equation}
\label{dm-3}
\rho_{m} = \xi_{2}(t^{\alpha}exp(\beta t))^{\xi_{1}-3}
\end{equation}
Here, $\xi_{2}$ is the constant of integration.
\begin{widetext}
\begin{equation}
\label{de-3}
\rho_{de} =	{\phi _0 (\delta  (3-\frac{\delta  \omega}{2})+3) (\frac{\alpha}{t} +\beta)^2 \left(t^{\alpha } e^{\beta  t}\right)^{\delta }}-\frac{k^2}{\phi _0\left( t^{\alpha } e^{\beta  t}\right)^{\delta +6}} -\xi _1 \left(t^{\alpha } e^{\beta  t}\right)^{\xi_{2} -3}
\end{equation}
\end{widetext}
The equation of state parameter of DE for interacting model is obtained as
\begin{widetext}
\begin{equation}
\label{omega}
\omega_{de} = -\frac{\frac{k^2}{\left(t^{\alpha } e^{\beta  t}\right)^{\delta +6}}+\phi _0^2 \left(t^{\alpha } e^{\beta  t}\right)^{\delta } \left(\left(\delta ^2 \left(\frac{\omega }{2}+1\right)+2 \delta +3\right) (\frac{\alpha}{t} +\beta)^2-\frac{\alpha  (\delta +2)}{t^2}\right)}{\phi^2 _0 (\delta  (3-\frac{\delta  \omega}{2})+3) (\frac{\alpha}{t} +\beta)^2 \left(t^{\alpha } e^{\beta  t}\right)^{\delta }-\frac{k^2}{\left( t^{\alpha } e^{\beta  t}\right)^{\delta +6}}-\phi _0 \xi _{1} \left(t^{\alpha } e^{\beta  t}\right)^{\xi_{2} -3}}
\end{equation}
\end{widetext}
It is worthwhile to note that we have not used Eqs. (\ref{ec-2}) and (\ref{ec-4}) for obtaining expression for $\rho_{de}$ and $\omega_{de}$ because one can not obtain explicit solutions of these equation for $\omega_{de} = \omega_{de}(t)$. The possible solution of these equations exist only when we choose either $\omega_{de} = constant$ or $\rho_{de}$ is in the form of special type DE density (Holographic \cite{Banerjee/2007}, Tsallis \cite{Aditya/2019} DE density). We used Eqs. (\ref{ef-H1}) and (\ref{ef-H2}) to get the expression for $\rho_{de}$ and $\omega_{de}$. So, our approach is different from other investigations in BD cosmology \cite{Banerjee/2007},\cite{Aditya/2019} and it is realistic because $\rho_{de}$ and $\omega_{de}$ vary with change in space-time curvature.
\section{The physical behavior of model and discussion}\label{6}
\begin{figure}\label{fig2}
\begin{center}
\includegraphics[width=4cm]{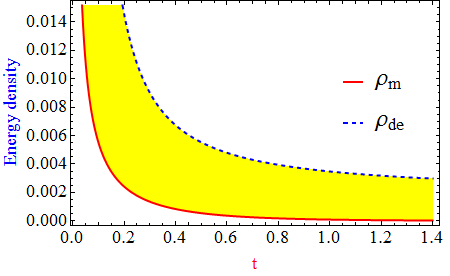}
\includegraphics[width=4cm]{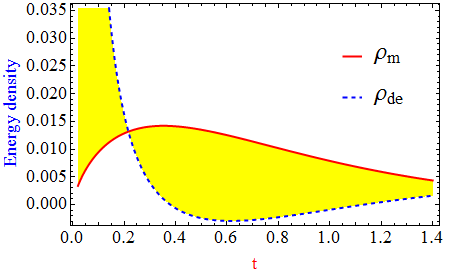}
\end{center}
\caption{The plot of energy densities of DM and DE versus time for non interacting model (left panel) and interacting model (right panel) for $\alpha = 0.281$ and $\beta = 0.796$.}
\end{figure}
\begin{figure}\label{fig2}
\begin{center}
\includegraphics[width=7cm]{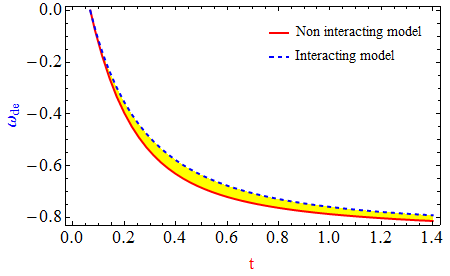}
\end{center}
\caption{The dynamics of equation of state parameter of DE versus time for $\alpha = 0.281$ and $\beta = 0.796$.}
\end{figure}
The graphical behaviors of energy densities of DM and DE for non interacting and variable coupling models is shown in Fig. 5. In non interacting scenario, $\rho_{m}$ and $\rho_{de}$ of model under consideration varies separately with time (see left panel of Fig. 5). From right panel of Fig. 5, we observe that the variation of $\rho_{m}$ and $\rho_{de}$ with respect to $t$ are coupled with one another. The variation of equation of state parameter of DE $(\omega_{de})$ versus time for non interacting, constant coupling and variable coupling model is depicted in Fig. 6. It has been seen that $\omega_{de}$ evolves with negative sign as time increases. The different DE models with negative $\omega_{de}$ were proposed instead of the constant vacuum energy density. Note that $\omega_{de} > -1$ and $\omega_{de} < -1$ are representing quintessence \cite{Steinhardt/1999} and phantom \cite{ref19} respectively. While $\omega_{de} = -1$ represents the cosmological constant dominated Universe and $\omega << -1$ is ruled out by SN Ia observations (Supernovae Legacy Survey, Gold sample of Hubble Spac Telescope \cite{Riess/2004}). Thus, the evolving range of $\omega_{de}$ of our derived model is in favor of quintessence Universe at present epoch. 
\subsection{Om(z) analysis}
The Om(z) parameter is a combination of the Hubble parameter $H$ and the redshift $z$. The Om(z) parameter of the universe in derived model is obtained as
\begin{equation}
Om(z)=\frac{\left[\frac{H(z)}{H_0}\right]^2-1}{(1+z)^3-1}
\end{equation}
where $H_0$ is the present Hubble parameter.\\
The negative, zero and positive values of Om$(z)$ parameter are used to differentiate DE models as the quintessence, $\Lambda$CDM and phantom DE models respectively \cite{Shahalam/2015}. In the present model, the om(z) parameter is obtained as
\begin{equation}
\label{om-1}
Om(z) = \frac{\left[\frac{\beta}{H_{0}}\left(\frac{1}{f(z)}+1\right)^{2}\right]-1}{(1+z)^{3}-1}
\end{equation}
\begin{figure}\label{fig2}
\begin{center}
\includegraphics[width=7cm]{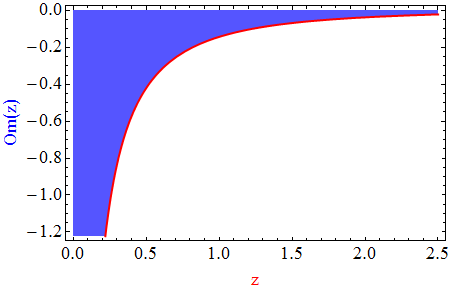}
\end{center}
\caption{The plot of om(z) versus redshift $z$ for $\alpha = 0.281$ and $\beta = 0.796$.}
\end{figure}
The variation of $om(z)$ parameter versus $z$ is depicted in Fig. 7. The $om(z)$ parameter is negative and monotonically decreases with decreasing values of $z$. Therefore, the derived model is describing a model of quintessence Universe.
\subsection{The shear scalar \& relative anisotropy}
\begin{figure}\label{fig2}
\begin{center}
\includegraphics[width=7cm]{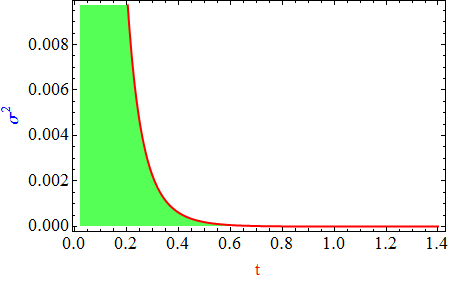}
\end{center}
\caption{The plot of $\sigma^{2}$ versus $t$ for $\alpha = 0.281$ and $\beta = 0.796$.}
\end{figure}
The shear scalar is obtained as
\begin{equation}
\label{ss}
\sigma^{2} = \frac{1}{2}\sigma_{ij}\sigma^{ij}
\end{equation}
where $\sigma_{ij} = u_{i;j}-\theta(g_{ij}-u_{i}u_{j})$\\

Thus, an expression for shear scalar is obtained as follow
\begin{equation}
\label{ss-1}
\sigma^{2} = \frac{\dot{D}^{2}}{D^{2}} = \frac{k^{2}}{a^{6}\phi^{2}}=\frac{k^{2}}{\phi_{0}^{2}\left(t^{\alpha}exp(\beta t)\right)^{2\delta + 6}}
\end{equation}
From equation (\ref{ss-1}), we observe that the shear scalar is decreasing function of time and finally it approaches to zero with passage of time. This behavior of $\sigma^{2}$ is clearly depicted in Fig. 8.

The relative anisotropy is given by
\begin{equation}
\label{ra}
A_{m} = \frac{\sigma^{2}}{\rho_{m}}
\end{equation}
Equation (\ref{ra}) exhibits that the relative anisotropy follows similar pattern as shear scalar. This means that  relative anisotropy is decreasing function of time.
\subsection{The age of the Universe}
\begin{figure}\label{fig2}
\begin{center}
\includegraphics[width=7cm]{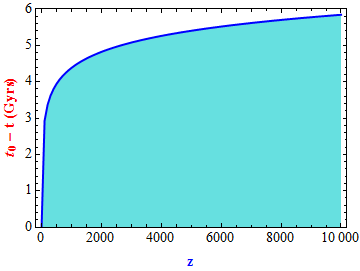}
\end{center}
\caption{The plot of $t_{0} - t$ versus $z$.}
\end{figure}
The age of the Universe is computed as
\begin{widetext}
\begin{align}
\label{age-1}
&dt = -\frac{dz}{(1+z)H(z)}\Rightarrow 
\int_{t}^{t_{0}} dt  =  \int_{0}^{z}\frac{1}{(1+z)H(z)}dz = \int_{0}^{z}\frac{1}{\beta(1+z)\left(\frac{1}{f(z)}+1\right)}dz 
\end{align}
\end{widetext}
Integrating equation (\ref{age-1}) numerically in the range $z \in \lbrace 0, 10^{10}\rbrace$ for $\alpha = 0.281\pm 0.004$ and $\beta = 0.796\pm 0.012$, we obtain the present age of the Universe in derived model is $14.61\pm 0.22$ Gyrs. Figure 9 depicts the age - redshift plot of the Universe in the model under consideration. From figure 9, we observe that at present $i. e.$ $z = 0$, $t = t_{0}$; $t_{0}$ is the present age of the Universe. It is important to note that the empirical value of age of the Universe in Plank collaboration results \cite{ref5} is obtained as $t_{0} = 13.81 \pm 0.038$ Gyrs. In some other cosmological investigations, age of the Universe is estimated as $14.46\pm 0.8$ Gyrs \cite{Bond/2013}, $14.3\pm0.6$ Gyrs \cite{Masi/2002} and $14.5\pm1.5$ Gyrs \cite{Renzini/1996}.
\section{Concluding remarks}\label{7}
In this paper, we have investigated an anisotropic model of transitioning Universe with hybrid scalar field in Brans-Dicke gravity. The Universe in derived model validates Mach's principle and have consistency with recent  observations. At $t = 0$, the average scale factor vanishes which is a simple consequence of the assumed functional form of $a(t)$. Also, we observe that the derived model starts with big bang singularity at $t = 0$. This singularity is point type as the average scale factor $a(t)$ vanishes at the initial moment $t = 0$. We have estimated the values of model parameters $\alpha$ and $\beta$ by bounding the Universe in derived model with recent OHD points. The best fit values of $\alpha$ and $\beta$ are obtained as $\alpha = 0.281\pm 0.004$ and $\beta = 0.796\pm 0.012$ at $68.3\%$ CL with $\chi^{2}_{min} = 24.178$ and  $\chi^{2}_{min}/dof = 0.549$. Here, dof stands for degree of freedom. The interacting and non-interacting scenarios of the Universe is investigated. Further, we observe that the eraly Universe was in decelerating phase while it is in accelerating phase of expansion at present (right panel of Fig. 4). The $om(z)$ analysis of derived model also shows that the present universe is dominated by dynamical dark energy fluid with effective equation of state parameter $\omega_{de} > -1$. We have depicted time and redshift evolution of various cosmological parameters and demonstrated the phase transition redshift through graphical representation (Fig. 4). It is found that the age of the Universe in derived model is $14.61\pm 0.22$ Gyrs at $68.3 \%$ CL which is $5.79 \%$ far from its empirical value obtained by Plank collaboration results \cite{ref5}. Further, we observe that the estimated age of the Universe in this paper has consistency with its value obtained in Refs \cite{Bond/2013,Masi/2002,Renzini/1996}. 
\section*{Acknowledgements}
The authors wish to place on record their sincere thanks to the honorable editor and referee for illuminating suggestions that have significantly improved our work in terms of research quality. The authors (AMA \& NA) express their gratitude to the Deanship of Scientific Research at King Khalid University for funding this work through the Research Group Program under grant number R.G.P. 2/25/40.\\
\section*{Appendix}
Eqs. (\ref{ef-H1}) and (\ref{ef-H2}) are re-written as
\begin{equation}
\label{ef-nH1}
2\frac{\ddot{a}}{a}+\frac{\dot{a}^{2}}{a^{2}} = -\frac{\omega_{de}\rho_{de}}{\phi} -\frac{k^{2}}{a^{6}\phi^{2}}-\frac{\omega}{2}\frac{\dot{\phi}^{2}}{\phi^{2}}-2\frac{\dot{\phi}}{\phi}\frac{\dot{a}}{a} - \frac{\ddot{\phi}}{\phi} = -p^{(eff)}
\end{equation}
\begin{equation}
\label{ef-nH2}
3\frac{\dot{a}^{2}}{a^{2}} =  \frac{\rho_{de}+\rho_{m}}{\phi} +\frac{k^{2}}{a^{6}\phi^{2}}+\frac{\omega}{2}\frac{\dot{\phi}^{2}}{\phi^{2}}-3\frac{\dot{\phi}}{\phi}\frac{\dot{a}}{a} = \rho^{(eff)}
\end{equation}
where $p^{(eff)} = p_{m}+p_{de}+p_{\sigma}+p_{\phi}$ and $\rho^{(eff)} = \rho_{m}+\rho_{de}+\rho_{\sigma}+\rho_{\phi}$. Note that $p_{\sigma}$, $p_{\phi}$, $\rho_{\sigma}$ and $\rho_{\phi}$ denote the pressures and energy densities due the anisotropy and BD scalar field \cite{Amirhashchi/2019,ref45}.\\
Differentiating Eq. (\ref{ef-nH2}) with respect to time, we can express the resulting answer as a linear combination of Eqs. (\ref{ef-nH1}) and (\ref{ef-nH2}). Therefore,
\begin{equation}
\label{ef-nH3}
\dot{\rho}^{(eff)}+3(\rho^{(eff)}+p^{(eff)})H = 0
\end{equation} 
The above equation is the direct consequence of the energy conservation law in BD cosmology.\\   

\end{document}